\documentclass[pra,aps,twocolumn,showpacs]{revtex4}
\usepackage{amsmath}
\usepackage{graphicx}
\usepackage{amssymb}

\begin{document}

\title{Efimov states near a Feshbach resonance}

\author{P. Massignan}

\author{H. T. C. Stoof}

\affiliation{Institute for Theoretical Physics, University of
Utrecht, Leuvenlaan 4, 3584 CE Utrecht, The Netherlands.}

\begin{abstract}
 We describe three-body collisions close to a Feshbach resonance by taking into account two-body scattering processes involving both the open and the closed channel. We extract the atom-dimer scattering length and the three-body recombination rate, predicting the existence at negative scattering length of a sharp minimum in the recombination losses due to the presence of a shallow bound level. We obtain very good agreement with the experimental results in atomic $^{133}$Cs of Kraemer \emph{et al.} [Nature \textbf{440}, 315 (2006)], and predict the position of Efimov resonances in a gas of $^{39}$K atoms.
\end{abstract}
\date{\today}
\maketitle

\section{Introduction}
 In the early seventies Efimov predicted that bosons with strong two-body interactions may bind together to form trimers with interesting universal properties, even under conditions where two-body bound states are not allowed \cite{Efimov70}. In particular, an infinite series of weakly bound three-body levels appears close to the continuum threshold when the magnitude of the \mbox{atom-atom} scattering length $|a|$ becomes large compared to the characteristic range of the potential $R$, which is of order $100a_0$ for alkali atoms ($a_0$ being the Bohr radius). In the region close to a resonance of $a$, these states are universal in the sense that they map onto each other via a discrete scaling symmetry, obtained by multiplying lengths and energies by a factor $\lambda$ and $1/\lambda^2$ respectively, where $\lambda\simeq 22.7$ \cite{BraatenReview06}. The observation of Efimov physics has been a major challenge in modern physics, due to the smallness of the typical binding energies. Experimental evidence of an Efimov state has been recently reported by Kraemer {\it et al.} in a thermal gas of $^{133}$Cs atoms \cite{Kraemer06}.
On the theoretical side, a microscopic description of Efimov physics is known to be particularly challenging. Indeed, the two-body scattering length $a$ alone is not sufficient to determine the solution of the three-body bosonic problem. Remarkable results have been obtained within the so-called "universal theory" (UT) \cite{BraatenReview06}, which assumes $|a| \gg R$ and a two-body T-matrix proportional to $(1+ika)^{-1}$, where $E=\hbar^2k^2/m$ is the energy in the center-of-mass frame of the two atoms. Nonetheless UT has \emph{a priori} no predictive power, since it contains two undetermined parameters, the first fixing the structure of the energy levels, and the second describing recombination into deeply-bound levels.

The analysis of current experiments in ultracold gases by means of UT might not properly work for different reasons. In first instance, experiments are usually performed outside the regime $|a|\gg R$, in order to avoid large three-body losses which in bosonic systems rise rapidly as $a^4$ close to a resonance. In addition, most of the interesting Feshbach resonances are characterized by background scattering lengths $|a_\mathrm{bg}|$ which are large compared to the range $R$, signaling the presence of a shallow (long-lived or bound) state in the open channel of the resonance which is not taken into account by the  scattering length approximation. It is therefore interesting to develop a model that does not rely on the assumptions of UT and compare it with the existing theories, such as those based on the hyper\-spherical method \cite{Stueck,hyperspherical,othersOnRecombRate} or the AGS approach \cite{Smirne07,Lee07}. 

\begin{figure}
\includegraphics[clip,width=\columnwidth]{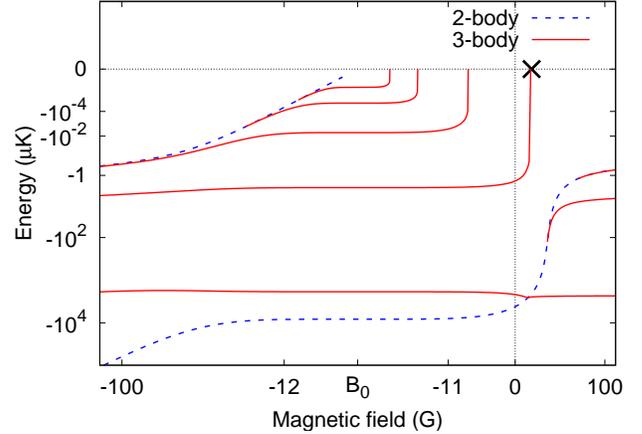}
\caption{(Color online) Two and three-body energy levels as a function of magnetic field for $^{133}$Cs. In order to visualize the Efimov states, both axes display highly nonlinear scales, obtained by plotting $(B-B_{0})^{1/5}$ and $E^{1/10}$. The cross shows the position of the recombination resonance observed in Ref.~\cite{Kraemer06}.
\label{rescaledEnergyLevels}}
\end{figure}

We choose here to work in momentum space, where the energy dependence of two-body interactions can be included in a natural way. We incorporate low-energy collisions between three identical bosons by means of a two-channel model that explicitly takes into account the role played by the shallow bound state in the open channel. 
We show that the presence of this bound state implies the possibility of having a sharp minimum in the three-body recombination even at negative scattering lengths, and it introduces noticeable modifications to the energy dependence of the low-lying trimer levels. 
As an example, we find that the trimer ground state in $^{133}$Cs never hits the three-body continuum threshold at large values of the magnetic field, as shown in Fig.~\ref{rescaledEnergyLevels}. We extract the temperature-dependent three-body recombination rate for $^{133}$Cs, finding very good agreement with the data presented in Ref.~\cite{Kraemer06}, and for $^{39}$K, a novel bosonic system  which appears to be a promising candidate for observing Efimov effects \cite{Roati07}. For both species we also present results for the atom-dimer scattering length, which may be of direct relevance to future experiments.

\section{Atom-dimer scattering}
\begin{figure}
\includegraphics[width=\columnwidth]{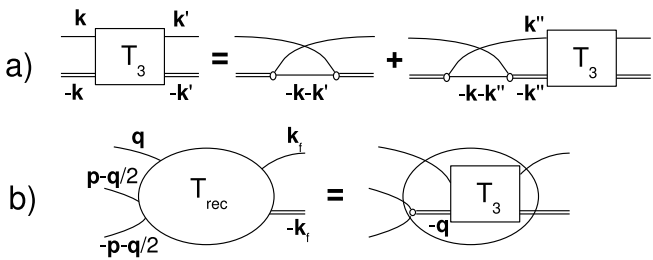}
\caption{Diagrams describing: a) atom-dimer scattering, and b)
recombination of three atoms into an atom and a dimer. \label{fig:Feynman-diagrams}}
\end{figure}

 In order to investigate three-body processes we introduce the transition matrix $T_3$, which describes the collision between an atom and a dimer. $T_3$ satisfies the Lippmann-Schwinger equation sketched in Fig.~\ref{fig:Feynman-diagrams}a.
 In the presence of purely s-wave interactions and provided that the two-body T-matrix $T_2(E)$ is analytic in the lower complex plane, the resulting integral equation reads
\cite{Skorniakov57, Levinsen06}
\begin{eqnarray}
\label{eq:STM}
T_{3}(k,k';E)=\frac{1}{2k k'}F(k') \hspace*{1.7in} \\
+\int_{0}^{\infty}\frac{d k''}{4\pi^{2}}
\frac{k''}{k}F(k'')T_{2}(E-3{k''}^{2}/4m) T_{3}(k'',k';E),
\nonumber
\end{eqnarray}
 where $E$ is the total energy in the center-of-mass reference frame, $k$ and $k'$ are the moduli of the momenta of the incoming and outgoing atoms and
$F(x)=\ln\left[(E-k^{2}-x^{2}+k x)/(E-k^{2}-x^{2}-k x)\right]$.
 The use of a simple $T_2$ proportional to $(1+ika)^{-1}$ in Eq.~(\ref{eq:STM}) presents a two-fold problem. In first instance, this form of the T-matrix does not carry any information on the presence of a shallow level in the open scattering channel, which plays a fundamental role in many experimental realizations of Feshbach physics. In addition, Eq.~(\ref{eq:STM}) with $T_2\propto(1+ika)^{-1}$ has a well-known ultraviolet cut-off dependence \cite{Danilov61}. In the following, we present a model which overcomes both problems.

The atom-atom scattering length near resonance can be written as
$a(B)=a_{\mathrm{bg}}(B)\left[1-\Delta{B}/(B-B_{0})\right]$, where
$B_{0}$ and $\Delta{B}$ are the location and the width of the resonance, and
$a_{\mathrm{bg}}(B)$ is the scattering length in the open channel
which may have a weak magnetic-field dependence \cite{aAAInKraemerEtAl}.
Physically, the processes involved in the dimer propagator $T_{2}(E)$ can be divided into collisions occurring in the open channel only, and collisions that proceed through the formation of a molecule in the closed channel \cite{Duine03,Bruun05}. The low-energy collisions between the atoms in the open channel contribute the term
$T_{\mathrm{bg}}(E)=T_{\mathrm{bg}}/\left[1-T_{\mathrm{bg}}\Xi(E)\right]$,
where $T_{\mathrm{bg}} \equiv T_{\mathrm{bg}}(0) =4\pi\hbar^{2}a_{\mathrm{bg}}/m$ and $\Xi(E)=m^{3/2}\sqrt{-E}/4\pi\hbar^{3}$ \cite{Duine03}.
The propagator for a molecule in the closed channel can be written
as $1/[E-\delta-\hbar\Sigma(E)]$, where
$\delta=\Delta\mu(B-B_{0})$ is the energy detuning from resonance,
and $\Delta\mu$ is the magnetic-moment difference between the open and closed channels. The
self-energy of the molecule is given by \cite{Duine03}
$\hbar\Sigma(E)=g^{2}\Xi(E)/\left[1-T_{\mathrm{bg}}\Xi(E)\right]$, where $g=\sqrt{T_{\mathrm{bg}}\Delta\mu\Delta B}$ is the atom-molecule coupling, and $g/\left[1-T_{\mathrm{bg}}\Xi(E)\right]$ corresponds to an atom-molecule vertex dressed by repeated scattering in the open channel. 
Finally, the T-matrix has to be multiplied by a factor of 2 to
take into account the possibility of interchanging the identical
incoming particles. By adding all the terms, one obtains
\cite{Bruun05,fermionicCase}
\begin{eqnarray}
\label{eq:T2} \frac{T_{2}(E)}{2} &=& T_{\mathrm{bg}}(E)
        +\left(\frac{g}{1-T_{\mathrm{bg}}\Xi(E)}\right)^{2}
               \frac{1}{E-\delta-\hbar\Sigma(E)}    \nonumber \\
&=& \frac{T_{\mathrm{bg}}}{\left(1+{\displaystyle
\frac{\Delta{\mu}\Delta{B}}{E-\delta}}\right)^{-1}
-T_{\mathrm{bg}}\Xi(E)}.
\end{eqnarray}
The pole structure of $T_{2}(E)$ quantitatively reproduces the low-energy Feshbach physics across the whole resonance.
In particular, if $a_{\mathrm{bg}}>0$ there are two branches, which describe the avoided crossing between the shallow bound states in the open and closed channel (see Fig.~\ref{rescaledEnergyLevels}).
If $a_{\mathrm{bg}}<0$ a single branch is present, in accord with the fact that the open channel no longer supports a shallow bound state. We have explicitly checked that the eigenstates of $T_2$ form an orthonormal set \cite{NoteOnOrthogonality}.

In order to ensure the convergence of Eq.~(\ref{eq:STM}), at large energies $T_2$ needs to fall-off at least as fast as $1/E$ \cite{Fedorov01}. In the absence of background interactions, i.e., if $a_\mathrm{bg}=0$, Eq.~(\ref{eq:T2}) would behave as $g^2/E$ and therefore would yield a convergent Eq.~(\ref{eq:STM}), but we see that by adding the background term we obtain $T_2(E)\propto 1/\sqrt{-E}$.
To restore the correct high-energy behaviour $g^2/E$, we add in the denominator of $T_2$ a term linear in energy that we write in the form $a_\mathrm{bg}R^*k^2$, where $R^*=\hbar^2/(m a_\mathrm{bg}\Delta B \Delta\mu)>0$ \cite{Petrov04}. In its final form, our two-body T-matrix reads
\begin{equation}
\label{eq:T2withRstar}
T_2(E)= \frac{2T_{\mathrm{bg}}}{\left(1+{\displaystyle
\frac{\Delta{\mu}\Delta{B}}{E-\delta}}\right)^{-1}
-T_{\mathrm{bg}}\Xi(E)+a_\mathrm{bg}R^*k^2}.
\end{equation}

 We assume here that only the two molecular states involved in the avoided crossing are important for the Efimov physics of interest, and that the effect of the much deeper lying molecular states can be neglected. Our T-matrix is expected to be accurate as long as $kR_0\ll1$, where $R_0=\sqrt{-\hbar^2/mE_0}\sim R$ is the length scale associated with the energy of the next lower lying state not included in the model. Since $R^*$ introduces in Eq.~(\ref{eq:STM}) an effective cut-off $k^*$ of order $1/R^*$, this amounts to require $R^*\gg R_0$. Our model is therefore ideally suited to describe three-body physics around narrow resonances with a large background scattering length, and has no free parameters. In case the condition $R^*\gg R_0$ should not be satisfied in the system under consideration, one may consider the quantity $R^*$ as a free parameter $R^*_{\mathrm{fit}}$, which plays then the role of an effective momentum cut-off. This single free parameter can then be adjusted to fit at best the available experimental data. This is to be compared to the 4 free parameters needed in the context of the universal one-channel model employed in Refs.~\cite{Kraemer06, othersOnRecombRate} to interpret the data at positive and negative values of $a$.

\section{Three-body recombination rate} We calculate here the rate at which three atoms recombine into an atom and a dimer with binding energy $E_\mathrm{b}$. This process is sketched in Fig.~\ref{fig:Feynman-diagrams}b. We work in the
center-of-mass frame of the three atoms and introduce the Jacobi
coordinates $\mathbf{p}$ and $\mathbf{q}$, in terms of which the
total incoming kinetic energy is diagonal and given by
$E=p^{2}/m+3q^{2}/4m$. The T-matrix describing the recombination
 depends only on the magnitude of $\mathbf{p}$ and $\mathbf{q}$, and is given in terms
of the dimer and trimer T-matrices by
$T_{\mathrm{rec}}(p,q;E_\mathrm{b})=3\sqrt{Z}T_{2}(p^{2}/m)T_{3}(k_{f},q;E)$
\cite{BraatenReview06}. The factor of 3 in this expression comes
from the sum of equivalent diagrams obtained by a cyclic
permutation of the three incoming bosons, and $k_f=\sqrt{4m(E-E_\mathrm{b})/3}$. The result
contains the square root of the residue $Z$ of the dimer propagator evaluated at its binding energy,
which takes into account the presence of one external dimer line.

\begin{figure}
\includegraphics[width=\columnwidth]{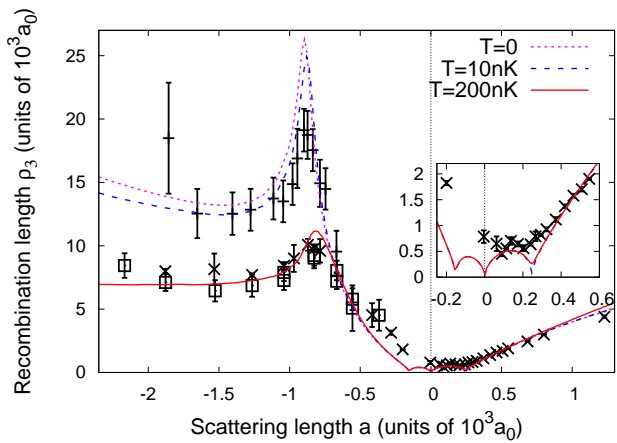}
\caption{(Color online) Three-body recombination length $\rho_3$ for $^{133}$Cs: the
curves represent our thermally-averaged results, while the
symbols are the experimental points of Ref. \cite{Kraemer06} at 10nK ($+$), 200nK ($\times$) and 250nK ($\square$).
\label{fig:Recombination-rate}}
\end{figure}

The three-body recombination rate is given by
Fermi's Golden Rule, which here yields
$\sum_{k_f} 2 k_{f}\left|T_{\mathrm{rec}}\right|^{2}/3\pi$. The sum over $k_f$ takes into account the fact that, when $a_{\mathrm{bg}}>0$ and $B<B_{0}$, one has to sum over the two possible outcoming molecular states with binding energies $E_\mathrm{b}^{+}$ and $E_\mathrm{b}^{-}$.
The variation in time of the atomic number density
$n$ is obtained by multiplying the rate by the number density of
triples of particles in the gas $n^{3}/3!$.
Assuming that after recombination both the atom and the dimer escape from the trap, we obtain $dn/dt=-3 L n^{3}$,
where the rate constant $L$ for a thermal gas is given by
\begin{equation}
L = \sum_{k_f} \left\langle \frac{k_{f}}{\pi}Z\left|T_{2}(p^{2}/m)T_{3}(k_{f},q;E)\right|^{2}
\right\rangle_{\rm th} ,
\label{rateConstantL}
\end{equation}
and $\langle \dots \rangle_{\rm th}$ denotes a thermal average
over the initial Jacobi momenta.

The recombination length $\rho_3=(6mL/\sqrt{3}\hbar)^{1/4}$ displays narrow peaks at the magnetic field values at which a trimer state hits the continuum of unbound atoms. Our results for $\rho_3$ are compared with the experimental data of Ref.~\cite{Kraemer06} in Fig.~\ref{fig:Recombination-rate}. Since for the broad cesium resonance $R^*=0.15a_0$ is not larger than the range $R$, we solve our model by keeping $R^*$ as a free parameter. We find that the best fit to the experimental data is obtained by choosing $R^*_{\mathrm{fit}}=0.14a_0$. The width and the height of the peak, as well as both wings at large values of $|a|$ are very well reproduced by our calculations at 10nK and 200nK. The recombination maximum at $a=-850a_0$ lowers and shifts towards larger values of magnetic field as the temperature is raised, in agreement with the fact that the Efimov state evolves into a triatomic resonance in the continuum. We also find a minimum of $\rho_{3}$ on the positive $a$ side at $a=240 a_0$, in agreement with the experiment.
 In addition, we observe a minimum of $\rho_{3}$ at a negative value of the scattering length, $a=-150a_0$. This minimum has the same origin as the ones present for $a>0$: it is due to destructive interference between two possible paths in the recombination process, the so-called St\"uckelberg oscillations \cite{Stueck}. Indeed, caesium possesses a shallow $6s$ two-body bound state even for $a<0$, and this state is precisely the $E_\mathrm{b}^-$ included in our theory (the lower blue dashed line in Fig.~\ref{rescaledEnergyLevels}). This novel non-universal prediction could be verified by a detailed measurement of the region $-200a_{0}<a<0$. We have checked that this minimum exists as long as $a_\mathrm{bg} \gg R$, i.e., as long as $E_\mathrm{b}^-$ falls within the range relevant to Efimov physics. We also note that our results for the height of the large recombination peak and the position of the minimum at $a>0$ are in much better agreement with experiment than the results of the parameter-free theory put forward by Lee \emph{et al.}~\cite{Lee07}.

One might wonder why $R^*_{\mathrm{fit}}$ is so close to $R^*$. The curious proximity of the two values may be explained by the fact that in $^{133}$Cs the next bound level lies deep down at $E_0=-125$MHz \cite{Chin04}, which implies $R_0=15a_0\ll R$. To extend the validity of our model when $R^*$ is not larger than $R_0$, we could in principle have included	 the effect of this and other deeper-lying bound
states by adding more poles to the two-body T-matrix $T_2(E)$. In Eq.~(\ref{eq:STM}) we have to integrate over these poles, which
physically has two effects. The real, or principle value, part of
the pole describes the virtual effects of the associated molecular
state, whereas the imaginary part describes the three-body
recombination processes forming this molecule. The integral over
the real part of the pole contains a large cancellation because of
the antisymmetric nature of the integrand. Moreover, recombination
processes forming deeply-bound molecular states are always
suppressed because of the strongly reduced overlap between the
initial and final states. Together, these observations indicate that sufficiently bound levels play a negligible role in the considered recombination process, and therefore they may partially explain the fact that $R^*_{\mathrm{fit}}\approx R^*$.

\begin{figure}
\includegraphics[width=\columnwidth]{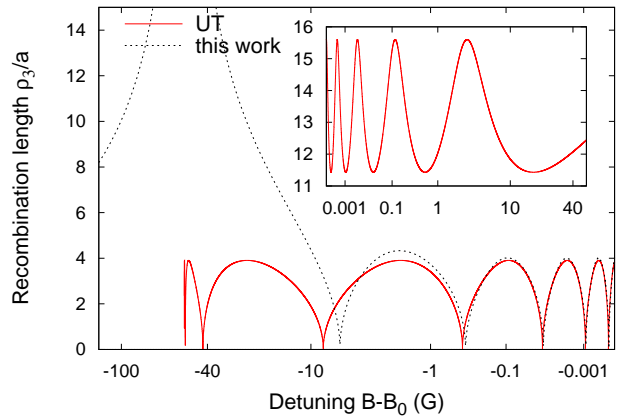}
\caption{(Color online) Recombination length $\rho_3/a$ at $T=0$ for $^{39}$K at negative detunings. The UT fit (with decay parameter $\eta=0$~\cite{BraatenReview06}) stops at $B-B_0=-52G$, where $a$ vanishes and changes sign. Inset: $\rho_3/a$ for positive detunings (UT fit with $a_*^\prime$ given by our model and $\eta=0.06$). 
\label{fig:recombRateK}}
\end{figure}

We now consider the three-body recombination for $^{39}$K atoms around the resonance located at $B_0=403$G. This resonance has a negative background scattering length $a_\mathrm{bg}=-29a_0$ and is characterized by $\delta\mu=1.5\mu_B$ \cite{D'Errico07}. Since for this resonance $R^*=29a_0$ is larger than $R_0=13a_0$, our model allows us to predict Efimov features without having to rely on existing experimental data. As shown in Fig.~\ref{fig:recombRateK}, non-universal effects play an important role far from the resonance, in a region that could easily be accessible experimentally. Since no shallow bound state exists for potassium at positive detuning $\delta$, Eq.~(\ref{rateConstantL}) may be used to evaluate $\rho_3$ only at $\delta<0$. Nonetheless, our approach allows us to extract the unknown phase of the universal result at $\delta>0$ by evaluating the scattering length $a_*^\prime$ at which one of the highly excited trimer levels hits the three-atom continuum.

\section{Atom-dimer scattering length}
\begin{figure}
\includegraphics[width=\columnwidth]{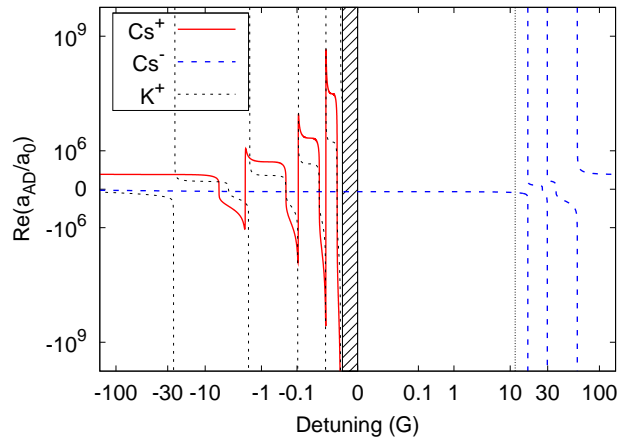}
\caption{(Color online) Real part of the atom-dimer scattering length: + (-) refers to collisions with the dimer of binding energy $E_\mathrm{b}^{+}$ ($E_\mathrm{b}^{-}$). The shaded region at $B\lesssim B_0$ contains a large number of resonances of $a_\mathrm{AD}^+$, each in correspondence with the appearance of a new Efimov state. The scales on the axes are obtained by plotting $(B-B_{0})^{1/5}$ and $a_\textrm{AD}^{1/5}$.
\label{fig:aAD}}
\end{figure}

We conclude the paper by discussing atom-dimer collisions. The atom-dimer scattering length $a_{\mathrm{AD}}$ is related to $T_{3}$ by
\begin{equation}
\frac{3\pi \hbar^{2} a_{\mathrm{AD}}}{m}=ZT_{3}(0,0;E_\mathrm{b}).
\label{eq:aAD}
\end{equation}
Here $2m/3$ is the atom-dimer reduced mass and the right hand-side is multiplied by a factor of $\sqrt Z$ for each of the two external dimer lines. The
atom-dimer scattering length is displayed in Fig.~\ref{fig:aAD}, and presents a resonance each time an Efimov state crosses the atom-dimer threshold.
In the case of caesium, $a_\mathrm{bg}>0$ and two dimers are present at $B<B_0$. As a result, the scattering length for the collision between an atom and the shallow dimer with energy $E=E_\mathrm{b}^+$ is complex, since this dimer may decay into the deeper one.
These resonances could in principle be observed by populating the molecular state (via RF-driven transitions, magnetic field sweeps, or three-body recombination) in $^{39}$K, where the Feshbach resonance is located at $B_0>0$.

We hope that our work will stimulate new experiments which are at present highly desirable to arrive at a complete microscopic understanding of the fascinating Efimov physics in ultracold Bose gases near a Feshbach resonance.

\section*{Acknowledgments}
 We wish to thank Mathijs Romans, Georg Bruun, Chris Pethick and Eric Braaten for valuable discussions, and Christoph N\"agerl for providing us with the experimental data of Fig.~\ref{fig:Recombination-rate}.
This work is supported by the Stichting voor Fundamenteel Onderzoek der Materie (FOM) and the Nederlandse Organisatie voor Wetenschaplijk Onderzoek (NWO).


\begin{thebibliography}{10}
\bibitem{aAAInKraemerEtAl}The Feshbach resonance employed in Ref.~\cite{Kraemer06} is characterized by
$\Delta{B}=28.72G$, $B_{0}=-11.74G$ and $a_{\mathrm{bg}}(B)=(1722+1.52 B/G)a_{0}$. 
We choose $\Delta\mu=2.68\mu_B$, in order to reproduce at $B=0$ the value $E_\mathrm{b}^-$=65MHz for the binding energy of the shallowest $^{133}$Cs$_2$ state measured in \cite{Chin04}.
\bibitem{fermionicCase}Equations (\ref{eq:STM}-\ref{eq:T2}) can be applied to the case of scattering between three identical fermions in two different spin states $(\uparrow \uparrow \downarrow)$ by removing the factor of 2 in Eq.~(\ref{eq:T2}) and by changing the sign of the r.h.s.~of Eq.~(\ref{eq:STM}).
\bibitem{NoteOnOrthogonality} The bound states are superpositions of bare-molecular and open-channel components, $|\psi^{\pm}\rangle = \sqrt{Z^{\pm}}|\chi_{m}\rangle\pm\sqrt{1-Z^{\pm}}|\chi^{\pm}\rangle$. Here $|\chi_{m}\rangle$ is the bare-molecular state in the closed channel, and $\chi^\pm(\mathbf{r})\propto\mathrm{e}^{-k_{\pm}r}/r$ is the normalized open-channel wave function with binding energy $E_{b}^{\pm}=-\hbar^{2}k_{\pm}^{2}/2m$. The residues are defined as $Z^{\pm}=\left(1-\hbar\ \partial_E\Sigma(E)|_{E=E_b^{\pm}}\right)^{-1}$.
\end{thebibliography}
\end{document}